\documentclass[twocolumn,amsmath,amssymb,aps,pra,superscriptaddress,showpacs,tightenlines,10pt,conference]{revtex4}
\usepackage{amsmath}
\usepackage{amsfonts}
\usepackage{graphicx}
\usepackage{epsfig}
\usepackage{color}
\usepackage{bm}
\usepackage[colorlinks,citecolor=blue,linkcolor=magenta]{hyperref}

\begin{document}

\title{Controllable nonlinearity in a dual-coupling optomechanical system under a weak-coupling regime }

\author{Gui-Lei Zhu}
\author{Xin-You L\"{u}}
\email{xinyoulu@hust.edu.cn}
\author{Liang-Liang Wan}
\author{Tai-Shuang Yin}
\author{Qian Bin}
\author{Ying Wu}
\email{yingwu2@126.com}
\affiliation{School of physics, Huazhong University of Science and Technology, Wuhan 430074, China}
\date{\today}
\begin{abstract}

Strong quantum nonlinearity gives rise to many interesting quantum effects and has wide applications in quantum physics. Here we investigate the quantum nonlinear effect of an optomechanical system (OMS) consisting of both linear and quadratic coupling. Interestingly, a controllable optomechanical nonlinearity is obtained by applying a driving laser into the cavity. This controllable optomechanical nonlinearity can be enhanced into a strong coupling regime, even if the system is initially in the weak coupling regime. Moreover, the system dissipation can be suppressed effectively, which allows the appearance of phonon sideband and photon blockade effects in the weak coupling regime. This work may inspire the exploration of dual-coupling optomechanical system as well as its applications in modern quantum science.

\end{abstract}
\maketitle
\section{Introduction}\label{sectionI}

Cavity optomechanics, exploring the nonlinear photon-phonon interaction via radiation pressure~\cite{Aspelmeyer2014,Xiong2015}, has achieved tremendous advances in recent years, including the realization of cooling a macroscopic mechanical resonator to ground state~\cite{Connell2010,Groblacher2009,Chan2011,Teufel2011}, optomechanically induced transparency~\cite{Weis2010,Safavi2011,Kronwald2013,Karuza2013}, coherent state conversion between cavity and mechanical modes~\cite{Fiore2011,Zhou2013,Palomaki2013}, and the generation of squeezed light~\cite{Brooks2012,Safavi2013,Purdy2013}. Those achievements offer the basic of exploring the applications of optomechanical nonlinearity which can be applied into quantum optics and quantum information sciences. In particular, recent studies have shown that the strong optomechanical nonlinearity could be used to generate single photon sources~\cite{Keller2004,Lounis2000,Rabl2011,Nunnenkamp2011}, engineer a nonclassical phonon state~\cite{Bose1997,Lund2004,Pepper2012,Lv2013}, and implement quantum information processing~\cite{Stannigel2012}.
However, the strong nonlinearity is difficult to realize in a normal optomechanical system due to its weak photon-phonon interaction~\cite{Aspelmeyer2014,Xiong2015}.

Recently, many methods have been proposed to enhance the radiation-pressure optomechanical coupling, such as using the photon hopping effects in two cavity systems~\cite{Stannigel2012,Komar2013,Ludwig2012} or multimode systems \cite{Massel2012,Seok2013,Buchmann2015,heinrich2011}, Josephson effect in superconducting circuits~\cite{Johansson2014,Heikkila2014,Rimberg2014}, and the optical~\cite{Lv20151,Lv20152} and mechanical parameter amplification \cite{Yin2017,Lemonde2016}. In the previous proposals, only the linear optomechanical coupling is considered and an additional subsystem is needed to be introduced into the OMS, which limits its applications in the optomechanical many-body lattices~\cite{Ludwig2013,Wan2017,Gan2016}.

\begin{figure}[t]
\centerline{\includegraphics[width=7.8cm]{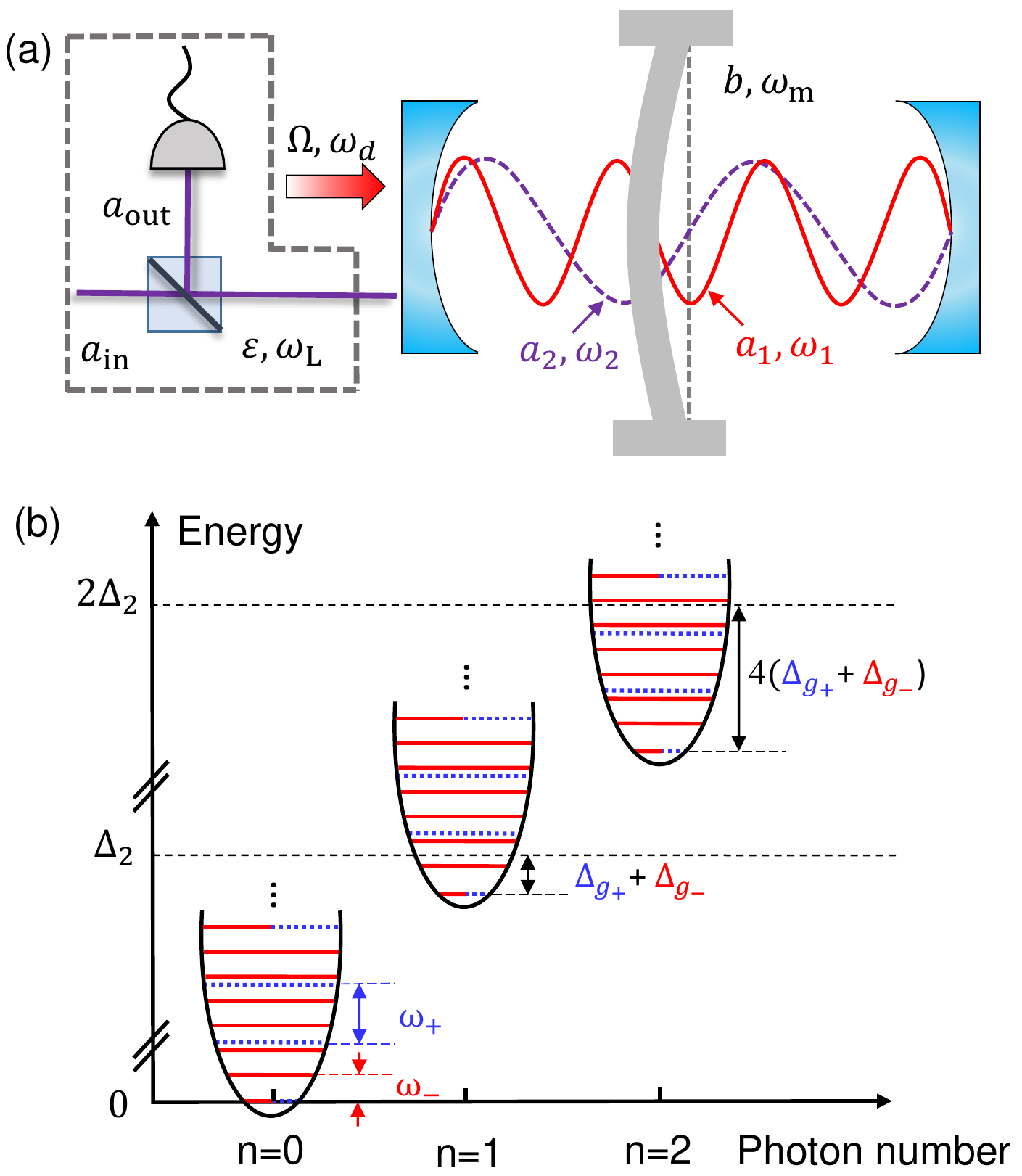}}
\caption{(Color online) (a) Schematic of a Fabry-Perot-type optomechanical system with a ``membrane-in-the-middle" configuration, consisting of two optical cavity modes $a_{1}$ and $a_{2}$ coupled to the mechanical mode $b$. The driving field on cavity mode $a_{1}$ is described by amplitude $\Omega$ and frequency $\omega_{\rm d}$. A weak probe field (enclosed by the dashed lines) with frequency $\omega_{\rm L}$ and amplitude $\varepsilon$ is applied to detect the nonlinear optomechanical system. (b) The eigenvalues of the effective Hamiltonian $H_{\rm OMS}$. Here, the red solid lines represent the energy level of $B_{-}$ mode while the blue dotted lines show the energy level of $B_{+}$. The parameters  $\Delta_{g_{\pm}}=g^{2}_{\pm}/\omega_{\pm}$ are the energy shift of polariton modes $B_{\pm}$.}
\label{fig1}
\end{figure}

Different from former works, here we investigate the optomechanical nonlinearity in an optomechanical system (OMS) with a membrane-in-the-middle configuration~\cite{Thompson2008,Sankey2010,Bhattacharya2008}, [see Fig.\,\ref{fig1}(a)], including both the linear and quadratic optomechanical coupling. Specifically, the mechanical mode oscillates around the antinode and node of the resonator mode $a_1$ and $a_2$, respectively. Typically, the quadratic coupling is much smaller than the linear coupling~\cite{Xuereb2013,Liao2013,Si2017,Vanner2011,Xin-You3}, which has been considered in the following discussion.

Interestingly,  we found that a controllable optomechanical nonlinearity could be obtained without introducing an additional subsystem. Physically, by strong driving the linear mode $a_1$, two effective polariton modes with controllable frequencies are obtained. They couple to the quadratic mode $a_2$ with the form of radiation pressure, and have the controllable interaction strengths and coupling weights. This ultimately leads to the results that the strong radiation-pressure coupling can be obtained in an initially weakly-coupled OMS. Moreover the present optomechanical nonlinearity can be controlled easily, i.e., adjusting the frequency or strength of the driving laser. To show this controllable nonlinearity, the controllable phonon sidebands and photon blockade effects are demonstrated in the weak coupling regime. This work is general and may also be applied into the photonic crystals~\cite{Krause2015,Eichenfield2009,Safavi-Naeini2010} and microtoroidal resonator~\cite{Vahala2003,Kippenberg2004,Hong2008}, even superconducting circuits with the ability of implementing quadratic interaction~\cite{Xiang2013,You2011,Kim2015}. It effectively associates the linear and quadratic optomechanical interactions in a same OMS, which is interesting in broadening the regimes of cavity optomechanics as well as its applications in modern quantum technologies.

This paper is organized as follows: In Sec.\,\ref{sectionII}, we introduce our model. The master equation describing the system's evolution and the controllable optomechanical nonlinearity of the proposed system are presented in Sec.\,\ref{sectionIII}. In Sec.\,\ref{sectionIV}, we discuss the nonlinear properties of this system featured by phonon sideband and photon blockade effects in the weak coupling regime. In Sec.\,\ref{sectionV}, we discuss the experimental prospect of our proposal. Finally, we conclude our results in Sec.\,\ref{sectionVI}. In addition, we provide a detailed derivation on the effective thermal occupancies in Appendix.

\section{The Model}\label{sectionII}
It is known that the optomechanical coupling between the optical cavity mode and the mechanical mode of an OMS depends strongly on the position of the middle mirror (or membrane)~\cite{Bhattacharya2008}. When the mirror is placed in the vicinity of the antinodes of one cavity mode, the maximum coupling between the cavity mode and mechanical mode is linear, in that quadratic coupling and higher order coupling terms are much smaller than the linear one and can be ignored safely. The other case is that the mirror is put around nodes of a cavity mode, the linear coupling is near to zero, accordingly, the maximum optomechanical coupling becomes quadratic.
\begin{figure}[t]
\centerline{\includegraphics[width=7.8cm]{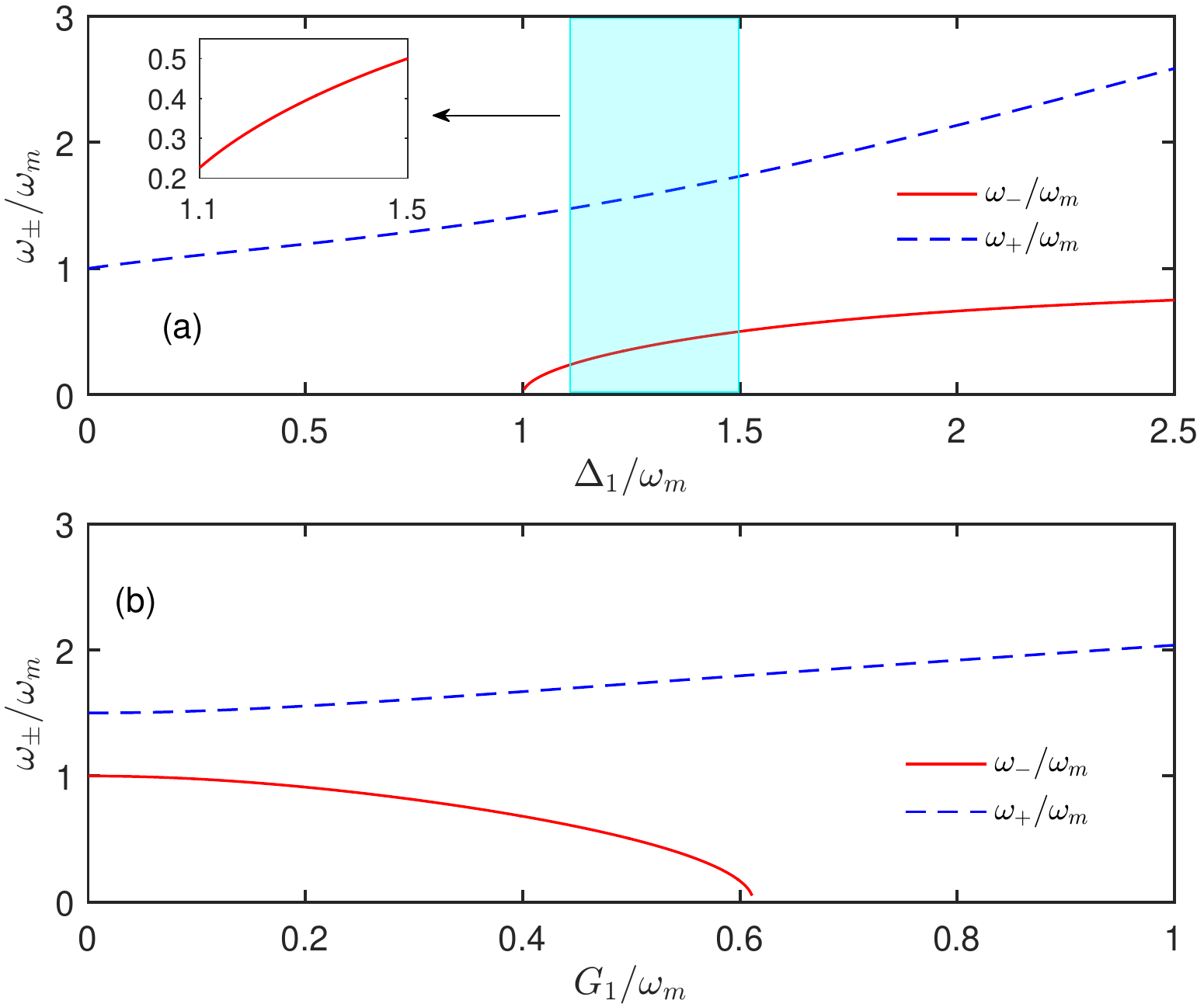}}
\caption{(Color online) The polariton modes $\omega_{\pm}$ of the optomechanical subsystem. Plots of the relationship between $\omega_{\pm}$ and (a) detuning $\Delta_{1}/\omega_{m}$, (b) the linearized coupling strength $G_{1}/\omega_{m}$. Main parameters including $\beta=25, g_{1}/\omega_{m}=10^{-2}$, in (b) $\delta_{1}/\omega_{m}=2.0$. The shadowed area indicates the considered parameter range in our proposal.}
\label{fig2}
\end{figure}

Given the above, here we consider a Fabry-Perot-type optomechanical cavity system with a ``membrane-in-the-middle" configuration, in which the mechanical mode (with frequency $\omega_{m}$) couples to both the cavity mode $a_1$ (with frequency $\omega_{1}$) linearly and $a_2$ (with frequency $\omega_{2}$) quadratically, as shown in Fig.\,\ref{fig1}(a). Specifically, the mechanical mode oscillates around the antinodes and nodes of mode $a_1$ and $a_2$, respectively. This dual-coupling optomechanical device is applicable to trap and cool the partially reflective mirrors driven by bichromatic lasers~\cite{Bhattacharya2008,Bhattacharya2007}.

In a frame rotating with frequency $\omega_{\rm d}$, the Hamiltonian of the optomechanical system depicted in Fig.\,\ref{fig1}(a) can be written as ($\hbar=1$)
\begin{eqnarray}
H_{\rm tot}&=&\delta_{1}a_{1}^{\dagger}a_{1}-g_{1}a_{1}^{\dagger}a_{1}(b+b^{\dagger})+\omega_{2}a_{2}^{\dagger}a_{2}\nonumber\\&&
-g_{2}a_{2}^{\dagger}a_{2}(b+b^{\dagger})^{2}+\omega_{m}b^{\dagger}b+i\Omega(a_{1}^{\dagger}-a_{1}),
\end{eqnarray}
where $a_{1} (a_{2})$ and $b$ are the annihilation operators of the optical cavity modes and the mechanical mode, and $g_{1} (g_{2})$ is the linear (quadratic) coupling strength between cavity mode $a_{1} (a_{2})$ and mechanical mode $b$. Here $\Omega$ represents the laser driving strength, $\delta_{1}=\omega_{1}-\omega_{\rm d}$ is the frequency detuning of optical cavity from the driving field. When a strong red-detuning driving field is applied, the cavity could generate large steady-state amplitudes in both the cavity and the mechanical modes. We assume $\alpha_{1}(\beta)$ is the steady-state amplitude of the cavity (mechanical) mode under the red-detuning driving. By using the displacement $a_{1}\to a_{1}+\alpha_{1},\, b\to b+\beta$, the system Hamiltonian $ H_{\rm tot}$ can be replaced by a shifted optomechanical Hamiltonian $H_{\rm shifted}=H_{\rm eff}+H_{\rm nl}$ given by
\begin{subequations}
\begin{eqnarray}
H_{\rm eff}&=&\Delta_{1}a_{1}^{\dagger}a_{1}+\Delta_{2} a_{2}^{\dagger}a_{2}-G_{2}a_{2}^{\dagger}a_{2}(b+b^{\dagger})\nonumber\\&&
-G_{1}(a_{1}^{\dagger}+a_{1})(b+b^{\dagger})+\omega_{m}b^{\dagger}b,   \\
H_{\rm nl}&=&-g_{1}a_{1}^{\dagger}a_{1}(b+b^{\dagger})-g_{2}a_{2}^{\dagger}a_{2}(b+b^{\dagger})^{2}.
\end{eqnarray}
\end{subequations}
 Here $G_{1}$ is the linearized optomechanical coupling strength, $G_2$ is the radiation-pressure coupling strength, and $\Delta_{1}$, $\Delta_{2}$ are the shifted detuning, given by
\begin{subequations}
\begin{align}
\Delta_{1}&=\delta_{1}-2g_{1}\beta,\, \Delta_{2}=\omega_{2}-4g_{2}\beta^{2}, \\
G_{1}&=g_{1}|\alpha_{1}|,\, G_{2}=4 g_{2}|\beta|.
\end{align}
\end{subequations}
The steady-state mean values $\alpha_{1} ,\beta$ satisfy $\omega_{m}\beta-g_{1}|\alpha_{1}|^{2}=0$. Without loss of generality, we have taken $\alpha_{1}, \beta,G_{1}, G_{2}$ to be real and positive.
Under the condition of $G_{1},G_{2}\gg g_{1},g_{2}$, the Hamiltonian $H_{\rm nl}$ can be ignored safely in our following calculations.

\begin{figure}[t]
\centerline{\includegraphics[width=8.0cm]{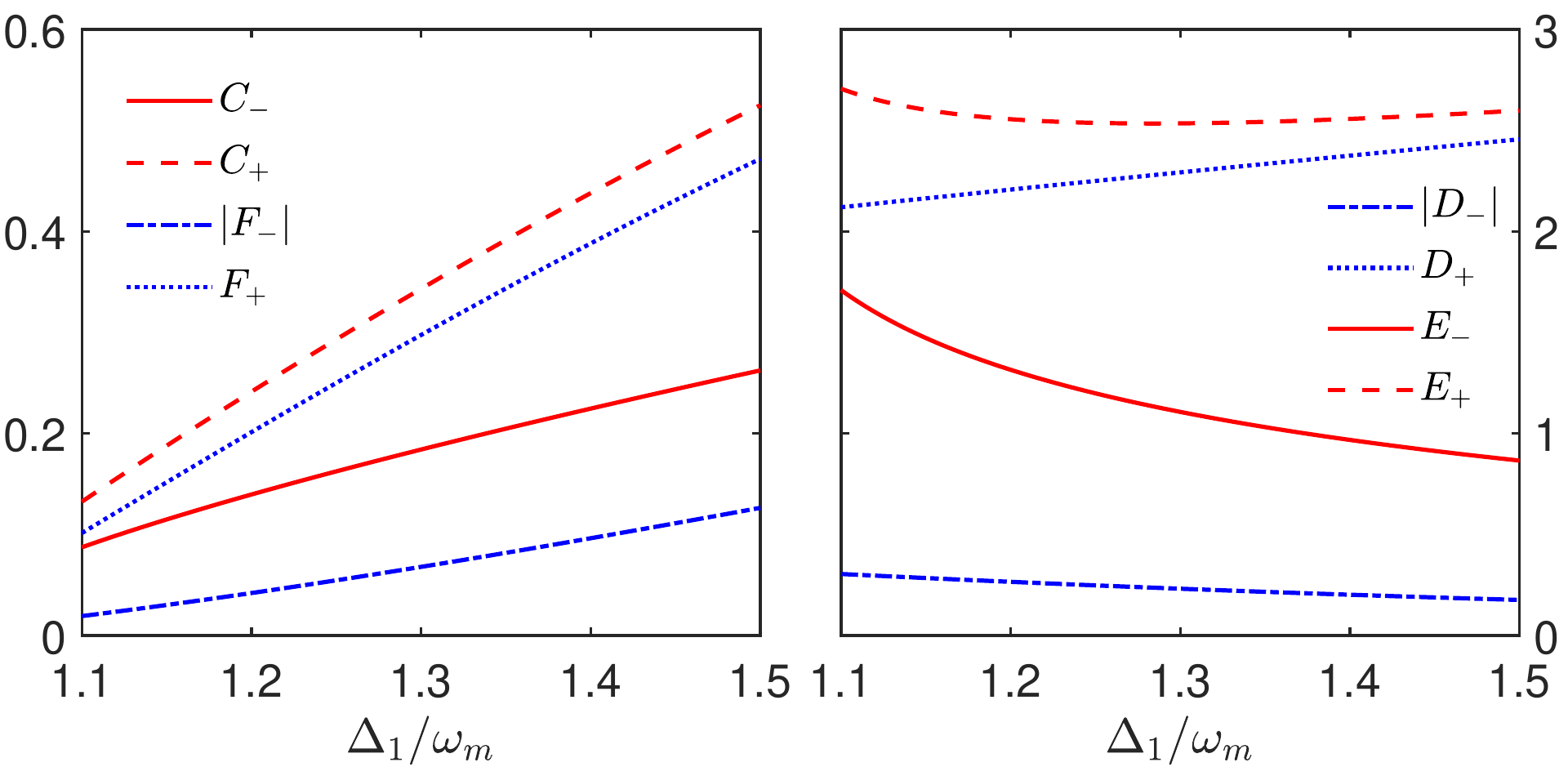}}
\caption{(Color online) The values of matrix factors $ C, |D|, E, |F|$ versus detuning $\Delta_{1}/\omega_{m}$ corresponding to the region of Fig.\,\ref{fig2}. The parameters are chosen as $\beta=25, g_{1}/\omega_{m}=10^{-2}$.}
\label{fig3}
\end{figure}

 From the Eq.\,(3b), we found that the introduced radiation-pressure coupling strength $G_2$ is proportional to the mechanical displacement $\beta$. In the following, we will give a simple analytical derivation to clarify this effect. Provided that the equilibrium position of membrane is effectively shifted with amplitude $\beta$, i.e., $b\to b+\beta$. Accordingly, we could obtain $g_{2}a_{2}^{\dagger}a_{2}(b+b^{\dagger}+2\beta)^{2}=g_{2}a_{2}^{\dagger}a_{2}(b+b^{\dagger})^2+G_2 a_{2}^{\dagger}a_{2}(b+b^{\dagger})+4\beta^2g_{2}a_{2}^{\dagger}a_{2}$. As is shown in this equation, firstly, the steady-state displacement of membrane can not enhance the originally quadratic coupling strength $g_2$. Secondly, a radiation-pressure coupling is introduced, and its strength $G_{2}$ is proportional to the displacement $\beta$. The last term $4\beta^2g_{2}a_{2}^{\dagger}a_{2}$ merely changes the cavity's resonant frequency.

\begin{figure}[t]
\centerline{\includegraphics[width=7.8cm]{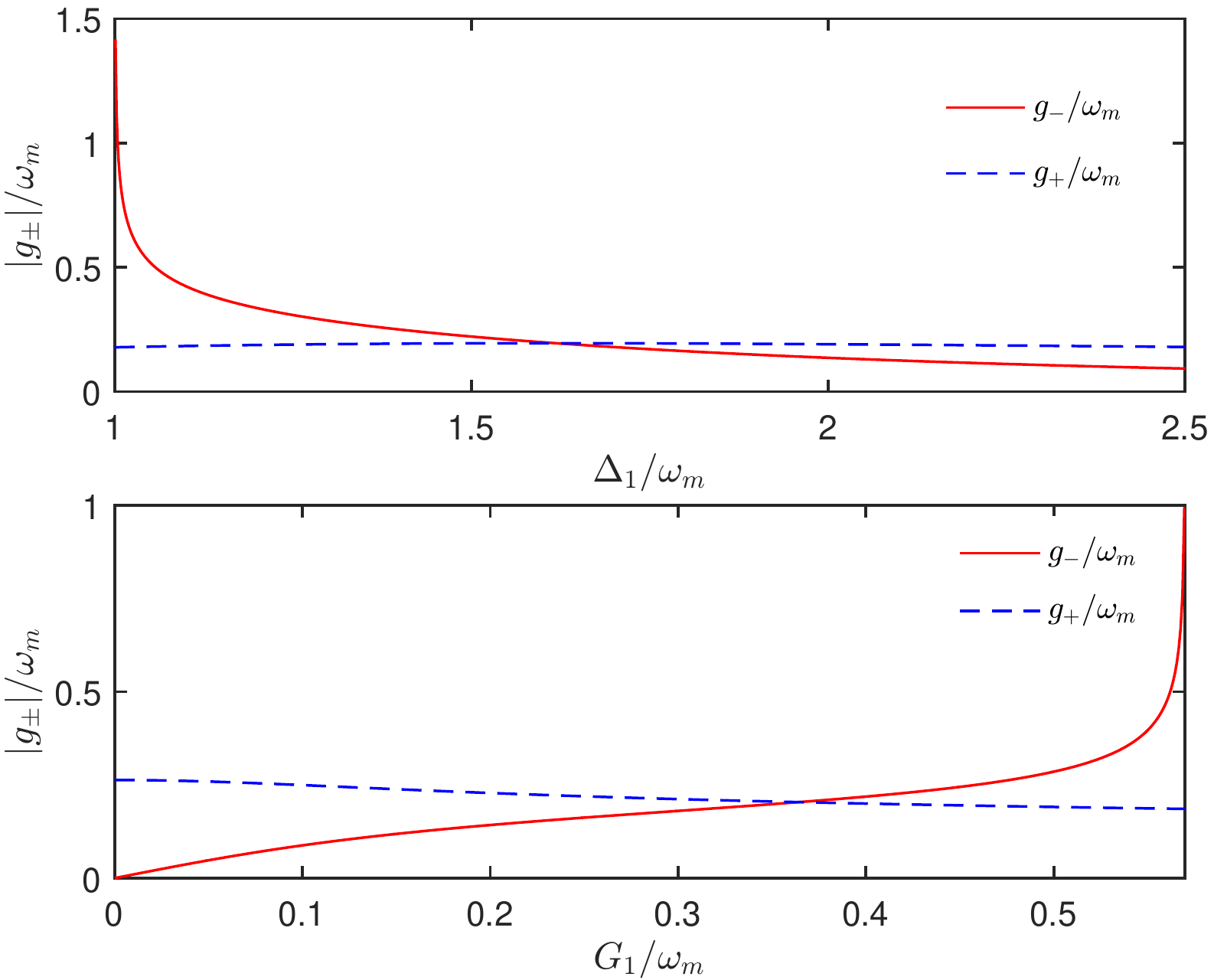}}
\caption{(Color online) The effective coupling strengths versus, (a) the detuning $\Delta_{1}/\omega_{m}$, (b) the linearized coupling strength $G_{1}/\omega_{m}$. The system parameters are $\delta_{1}/\omega_{m}=1.80$ for (a), other parameters are chosen as $\beta=25, g_{1}/\omega_{m}=10^{-2}, g_{2}/\omega_{m}=3\times10^{-3}$.}
\label{fig4}
\end{figure}

Physically, we could understand this effect as following. In our proposal, a strong driving field applied to cavity mode $a_{1}$ changes the effective equilibrium position of membrane, i.e., the membrane deviates from the nodes of $a_{2}$ mode. This shift introduces a radiation-pressure optomechanical coupling from the original quadratic optomechanical interaction. Then the dominated optomechanical coupling shifts from quadratic form [i.e., four-operators term $a_{2}^{\dagger}a_{2}(b+b^{\dagger})^{2}$] to a radiation-pressure form [i.e., three-operators term $a_{2}^{\dagger}a_{2}(b+b^{\dagger})$], in that the strength of the introduced radiation-pressure coupling is enhanced by the displacement amplitude $\beta$ of membrane's equilibrium position. Actually, this process is similar to the linearization procedure~\cite{Pace1993,Vitali2007,Wilson2007,Marquardt2007}, i.e., when a strong driving field is applied to a normal OMS with radiation-pressure coupling $g_{0}a^{\dagger}a(b+b^{\dagger})$, the dominated optomechanical interaction will shift from the radiation-pressure coupling to a linearized optomechanical interaction $G(a+a^{\dagger})(b+b^{\dagger})$. Because the linearized optomechanical coupling strength $G=g_{0}\alpha$ ($|\alpha|^2$ is the mean photon number in the cavity) can be much larger than $g_0$.

\section{Controllable optomechanical nonlinearity}\label{sectionIII}
Interestingly, the present system has a controllable optomechanical interaction, which leads to the controllable optomechanical nonlinearity.
Specifically, the Eq.\,(2a) can be diagonalized via a Bogoliubov transformation $R = M B$ with $R^{\rm T}=(a_{1},a_{1}^{\dagger},b,b^{\dagger})$ and $B^{\rm T}=(B_{-},B_{-}^{\dagger},B_{+},B_{+}^{\dagger})$. We refer $B_{\pm}$ as polariton modes including both photonic and phononic components, and the transformation matrix is given by \cite{Lv2013}
\begin{equation}
M=
\left(
\begin{array}{cccc}
C_{+}&C_{-}&D_{+}&D_{-} \\
C_{-}&C_{+}&D_{-}&D_{+}\\
-E_{+}&-E_{-}&F_{+}&F_{-} \\
-E_{-}&-E_{+}&F_{-}&F_{+}
\end{array}
\right),
\end{equation}
where the matrix factors $C_{\pm},D_{\pm},E_{\pm},F_{\pm}$ read,
\begin{subequations}
\begin{align}
C_{\pm}&=\frac{\rm cos \theta}{2}\frac{(\Delta_{1}\pm\omega_{-})}{\sqrt{\Delta_{1}\omega_{-}}}\,\,,\,\,\,\,\,\,
D_{\pm}=\frac{\rm sin \theta}{2}\frac{(\Delta_{1}\pm\omega_{+})}{\sqrt{\Delta_{1}\omega_{+}}} ,     \\
E_{\pm}&=\frac{\rm sin \theta}{2}\frac{(\omega_{m}\pm\omega_{-})}{\sqrt{\omega_{m}\omega_{-}}}\,\,,\,\,\,\,\,\,
F_{\pm}=\frac{\rm cos \theta}{2}\frac{(\omega_{m}\pm\omega_{+})}{\sqrt{\omega_{m}\omega_{+}}}.
\end{align}
\end{subequations}
 By using the inverse of matrix $M$, we obtain the expression of new mode $B_{\pm}$ in terms of $a_{1}$ and $b$ modes, i.e.,
 \begin{figure}[t]
\centerline{\includegraphics[width=7.8cm]{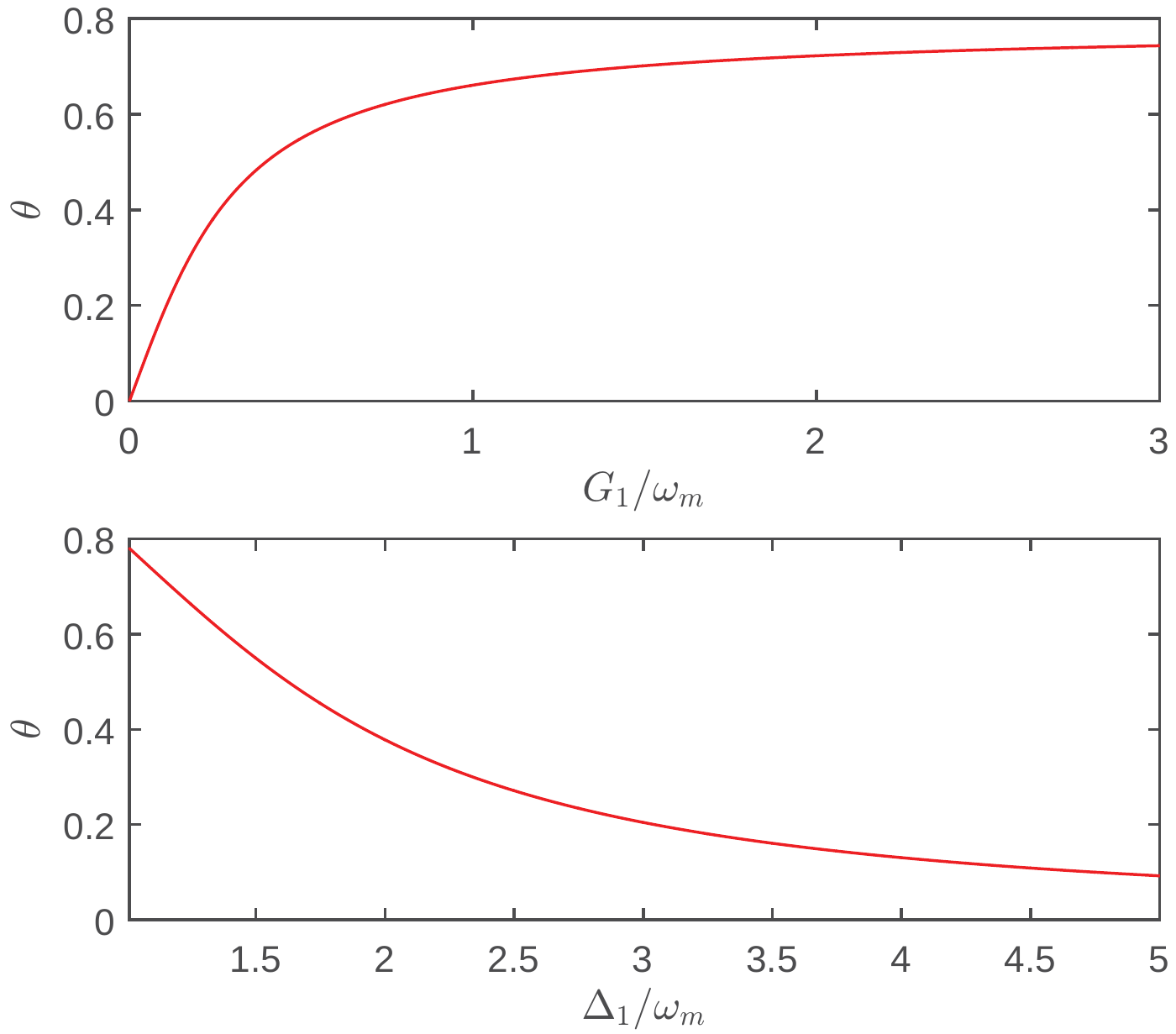}}
\caption{(Color online) Relationship between angle $\theta$ and linearized coupling strength $G_{1}/\omega_{m}$ and the detuning $\Delta_{1}/\omega_{m}$. For the parameters are $\beta=25, g_{1}/\omega_{m}=10^{-2} $ and $ \delta_{1}/\omega_{m}=2.0.$}
\label{fig5}
\end{figure}
\begin{subequations}
\begin{align}
B_{-}&=C_{+}a_{1}-C_{-}a_{1}^{\dagger}-E_{+}b+E_{-}b^{\dagger},  \\
B_{-}^{\dagger}&=-C_{-}a_{1}+C_{+}a_{1}^{\dagger}+E_{-}b-E_{+}b^{\dagger},  \\
B_{+}&=D_{+}a_{1}-D_{-}a_{1}^{\dagger}+F_{+}b-F_{-}b^{\dagger},  \\
B_{+}^{\dagger}&=-D_{-}a_{1}+D_{+}a_{1}^{\dagger}-F_{-}b+F_{+}b^{\dagger}.
\end{align}
\end{subequations}After a Bogoliubov transformation, we obtain a standard-like optomechanical Hamiltonian, given by
 \begin{eqnarray}
H_{\rm OMS}&=&\Delta_{2}a_{2}^{\dagger}a_{2}+\omega_{+}B_{+}^{\dagger}B_{+}-g_{+}a_{2}^{\dagger}a_{2}(B_{+}+B_{+}^{\dagger})\nonumber\\&&
+\omega_{-}B_{-}^{\dagger}B_{-}-g_{-}a_{2}^{\dagger}a_{2}(B_{-}+B_{-}^{\dagger}).
\end{eqnarray}
Here $\omega_{\pm}$ are the polariton mode frequencies of the subsystem,

\begin{eqnarray}
\omega_{\pm}^{2}&=&\frac{1}{2}(\Delta_{1}^{2}+\omega_{m}^{2}\pm\sqrt{(\omega_{m}^{2}-\Delta_{1}^{2})^{2}+
16G_{1}^{2}\Delta_{1}\omega_{m}})\,\,\,\,\,\,\,\,\,\,\,\,
\end{eqnarray}
and the effective coupling strengths of the optomechanical subsystem are given by
\begin{subequations}
\begin{align}
g_{-}=&-4g_{2}\beta \rm sin \theta \sqrt{\frac{\omega_{m}}{\omega_{-}}} , \\
g_{+}=&+4g_{2}\beta \rm cos \theta \sqrt{\frac{\omega_{m}}{\omega_{+}}} ,
\end{align}
\end{subequations}
with the angle $\theta$ defined as
\begin{align}
{\rm tan\,} 2\theta =\frac{4G_{1}\sqrt{\Delta_{1}\omega_{m}}}{\Delta_{1}^{2}-\omega_{m}^{2}}.
\end{align}
\begin{figure}[t]
\centerline{\includegraphics[width=8cm]{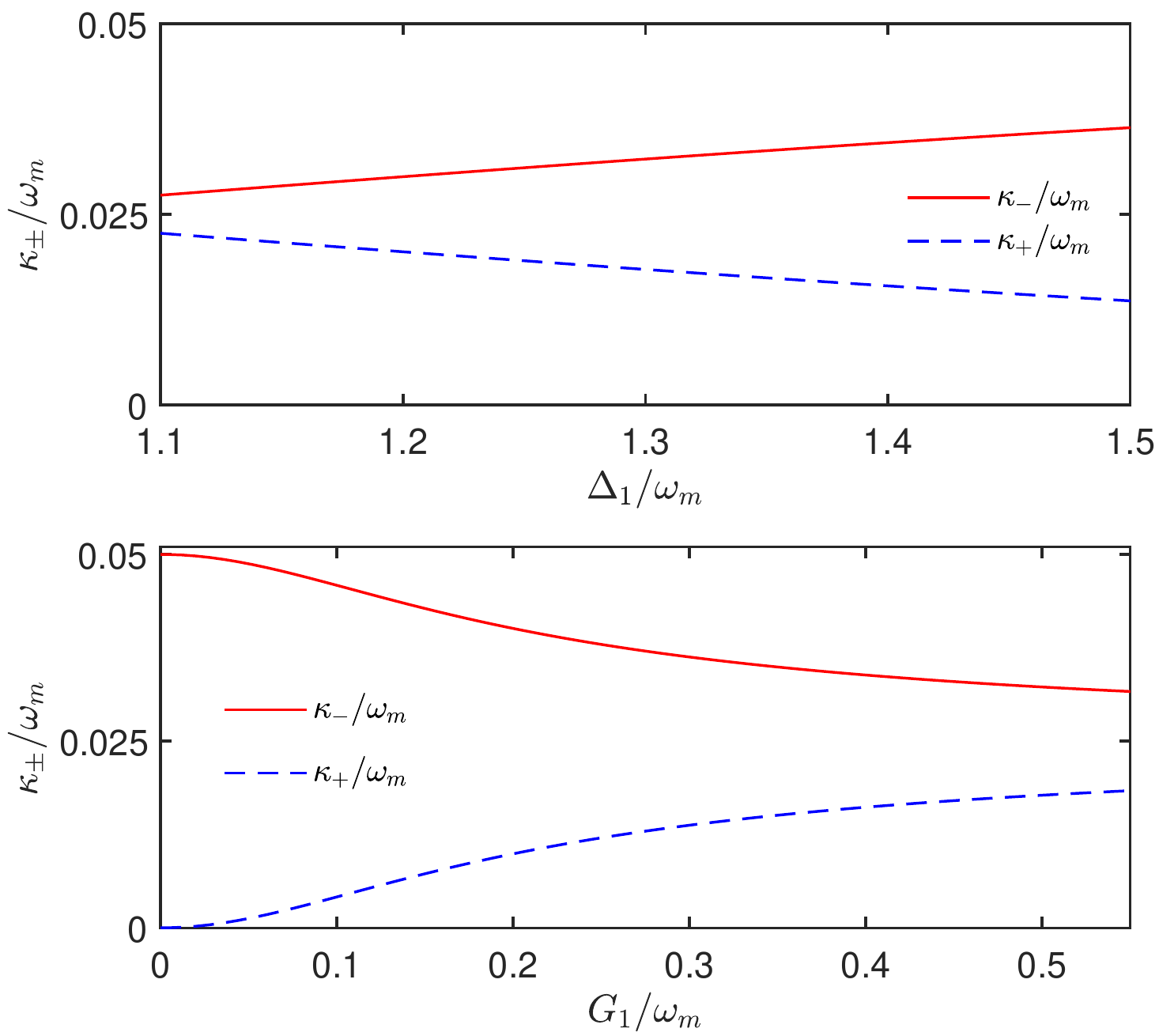}}
\caption{(Color online) The corresponding effective damping rates of subsystem. Main parameters are $\beta=25,g_{1}/\omega_{m}=10^{-2},\gamma/\omega_{m}=10^{-5},\kappa/\omega_{m}=5\times10^{-2}$, and
$\delta_{1}/\omega_{m}=1.80$ for (b).}
\label{fig6}
\end{figure}
From the Eqs.\,(6), we could obtain polariton modes $B_{\pm}$ of optical and mechanical modes. As discussed in Sec.\,\ref{sectionII}, the introduced radiation-pressure coupling strength $G_2$ could be enhanced by the displacement amplitude $\beta$.  Moreover, as is shown in the Eq.(9a), the effective subsystem's radiation-pressure coupling strength $g_{-}$ can be enhanced when the polariton mode frequency $\omega_{-}$ is decreased. Physically, this result can be understood as following. When the polariton mode frequency $\omega_{-}$ decreases, the polariton mode $B_{-}$ coupled to the optical field is highly softened. Accordingly, the subsystem's radiation-pressure coupling $g_{-}$ is enhanced largely. In addition, in the extreme case of $\omega_{-}\rightarrow0$, $g_{-}$ approaches to infinity. However, in our proposal, the considered region is far from $\omega_-=0$, which avoids the appearance of this singular point.

Specifically, equation (8) shows that $G_{1}=\sqrt{\Delta_{1}\omega_{m}}/2$ corresponds to a critical point of $\omega_{-}^{2}=0$.
Theoretically, when $\omega_{-}\to 0$, the matrix factors $C, D, E, F$ approach to infinity, then the new modes $B_{\pm}$ tend to be zero (zero combinations of mode $a_{1}$ and $b$). However, in our proposal, we have taken the value of $\Delta_{1}$ from $1.1\omega_{m}$ to $1.5\omega_{m}$, accordingly $\omega_{-}$ ranges from $0.25\omega_{m}$ to $0.5\omega_{m}$. As shown in the insert of Fig.\,\ref{fig2}(a), the chosen parameter region of $\omega_{-}$ is far away from the critical point $\omega_{-}=0$. In order to make it more explicit, we plotted the function of matrix factors $C, D, E, F$ versus detuning $\Delta_{1}/\omega_{m}$ in Fig.\,\ref{fig3}. It is shown that, in the considered parameter region, the matrix factors $C, D, E, F$ are finite, which ensures the validity of polariton mode $B_{-}$.

In Fig.\,\ref{fig4}, we present the dependence of the effective coupling strength $g_{\pm}$ on system parameters $\Delta_1$ and $G_1$. It is shown that, in the considered region, $g_{-}$ can be enhanced. Together with the validity of the polariton mode $B_{-}$, the enhancement of system nonlinearity decided by increasing $g_{-}$ has real meaning.
Moreover, here the optical mode couples to the two effective polariton modes with a weight decided by $\theta$. As shown in Fig.\,\ref{fig5}, $\theta$ ranges from 0 to 0.75 and also can be controlled by modulating the detuning $\Delta_{1}/\omega_{m}$. By adjusting the driving laser applied into mode $a_1$, the proposed system has a controllable optomechanical nonlinearity decided by the controllable optomechanical interactions.
\begin{figure}[t]
\centerline{\includegraphics[width=7.8cm]{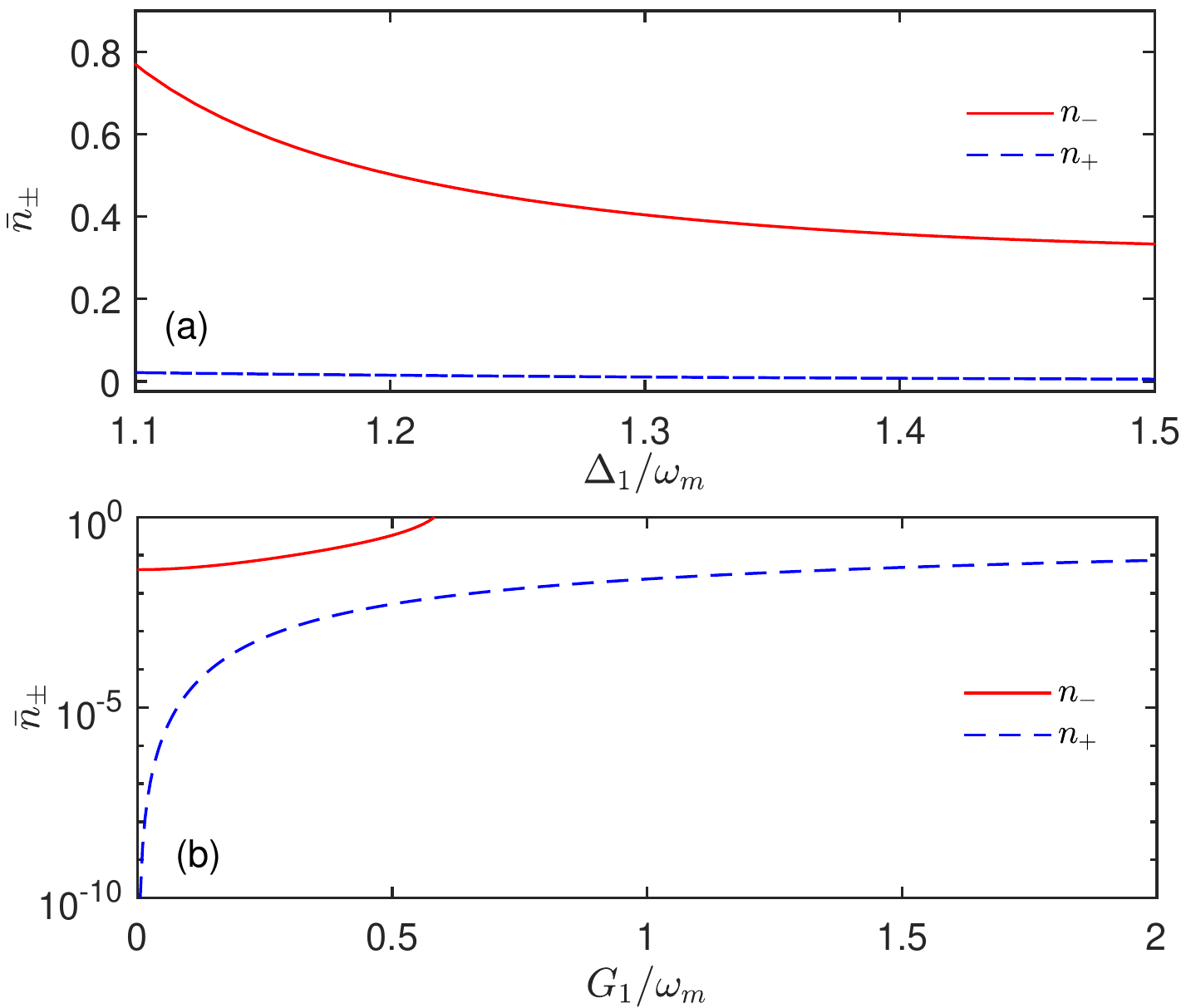}}
\caption{(Color online) The resulting effective bath thermal occupancies of the cavity and mechanical modes. Parameters are $\beta=25, g_{1}/\omega_{m}=10^{-2}, \gamma/\omega_{m}=10^{-5},\kappa/\omega_{m}=5\times10^{-2}$, and $\delta_{1}/\omega_{m}=2.0$ for (b).}
\label{fig7}
\end{figure}

\section{Phonon sideband and photon blockade effect in the weak coupling regime}\label{sectionIV}
Generally speaking, the strong optomechanical nonlinearity can induce phonon sideband in the excitation spectrum and photon blockade under the condition of a weak driving field. In order to probe (or utilize) this controllable nonlinearity, we drive the cavity mode $a_{2}$ using a weak probe field with frequency $\omega_{\rm L}$, amplitude $\varepsilon \,(\varepsilon\ll \kappa)$, where $\kappa$ is the initial cavity damping rate. Then the effective system Hamiltonian becomes

\begin{align}
H_{\rm OMS}^{'}=&\Delta_{2}^{'}a_{2}^{\dagger}a_{2}+\sum_{\sigma=\pm}\omega_{\sigma}B_{\sigma}^{\dagger}B_{\sigma}
-\sum_{\sigma=\pm}g_{\sigma}a_{2}^{\dagger}a_{2}(B_{\sigma}+B_{\sigma}^{\dagger}) \nonumber \\ +&i\varepsilon(a_{2}^{\dagger}-a_{2}),
\end{align}
where $\Delta_{2}^{'}=\Delta_{2}-\omega_{\rm L}$. By diagonalizing $H^{'}_{\rm OMS}$ with the transformation $H^{'}_{\rm OMS}\to UH^{'}_{\rm OMS}U$ and $U=e^{-iPa_{2}^{\dagger}a_{2}}$, $P=i\sum_{\sigma}(\frac{g_{\sigma}}{\omega_{\sigma}})(B_{\sigma}^{\dagger}-B_{\sigma})$, we can obtain
\begin{align}
H_{\rm OMS}^{'}=&\Delta_{2}^{'}a_{2}^{\dagger}a_{2}+\sum_{\sigma}(\omega_{\sigma}B_{\sigma}^{\dagger}B_{\sigma}-
\frac{g_{\sigma}^{2}}{\omega_{\sigma}}a_{2}^{\dagger}a_{2}^{\dagger}a_{2}a_{2})  \nonumber \\
+&i\varepsilon(a_{2}^{\dagger}e^{iP}-e^{-iP}a_{2}).
\end{align}
Then the system eigenstates are $|n,\tilde{m}_{+},\tilde{m}_{-}\rangle$ and corresponding eigenvalues given by
\begin{align}
E_{n,\tilde{m}_{\sigma}} =n\Delta_{2}-\sum_{\sigma}(n^{2}\frac{g_{\sigma}^{2}}{\omega_{\sigma}}-m_{\sigma}
\omega_{\sigma}),
\end{align}
where $n,m_{\sigma}$ are the non-negative integers and $|n,\tilde{m}_{+}, \tilde m_{-}\rangle =U |n,m_{+}, m_{-}\rangle$. Here $|n,m_{+}, m_{-}\rangle$ represents a state of $n$ photons and $m_{\sigma}$ polaritons.

 The results mentioned above show that we obtain an optomechanical system with two effective polariton modes (i.e., $B_{\pm}$) coupled to a same cavity mode $a_{2}$. The corresponding coupling weight decided by $\theta$ can be controlled by tuning the frequency detuning $\delta_{1}$. Qualitatively, we present the energy level structure of system in Fig.\,\ref{fig1}(b), which clearly shows the strong Kerr nonlinearity of system coming from both of the two optomechanical interactions between $a_2$ and $B_{\pm}$. Then, to show quantitatively the nonlinearity induced phonon sideband and photon blockade, we numerically calculate system dynamics including the cavity mode $a_2$ and two polariton modes $B_{\pm}$.

\begin{figure}[t]
\centerline{\includegraphics[width=7.8cm]{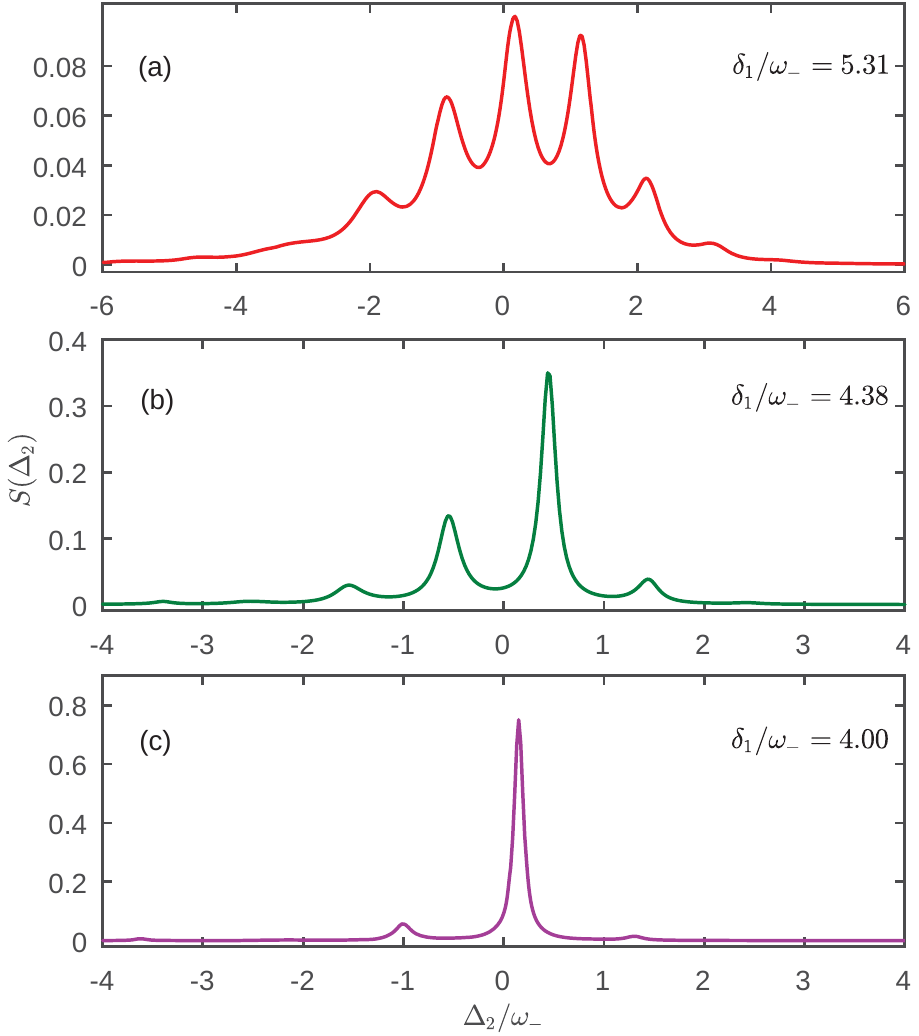}}
\caption{(Color online) Optical cavity output spectrum $S(\Delta_{2})$ for three different values of the detuning $\delta_{1}/\omega_{-}$. (a) $\delta_{1}/\omega_{-}=5.31$, $g_{1}/\omega_{-}=3.00\times10^{-2}$, $g_{2}/\omega_{-}=9.36\times10^{-3}$ and (b) $\delta_{1}/\omega_{-}=4.38$, $g_{1}/\omega_{-}=2.37\times10^{-2}$, $g_{2}/\omega_{-}=7.10\times10^{-3}$ and (c) $\delta_{1}/\omega_{-}=4.00$, $g_{1}/\omega_{-}=2.00\times10^{-2}$, $g_{2}/\omega_{-}=6.00\times10^{-3}$.}
\label{fig8}
\end{figure}
Particularly, within a Lindblad approach for the system dissipation, the system dynamics is decided by the master equation given by~\cite{Lemonde2015}
\begin{figure}[t]
\centerline{\includegraphics[width=7.8cm]{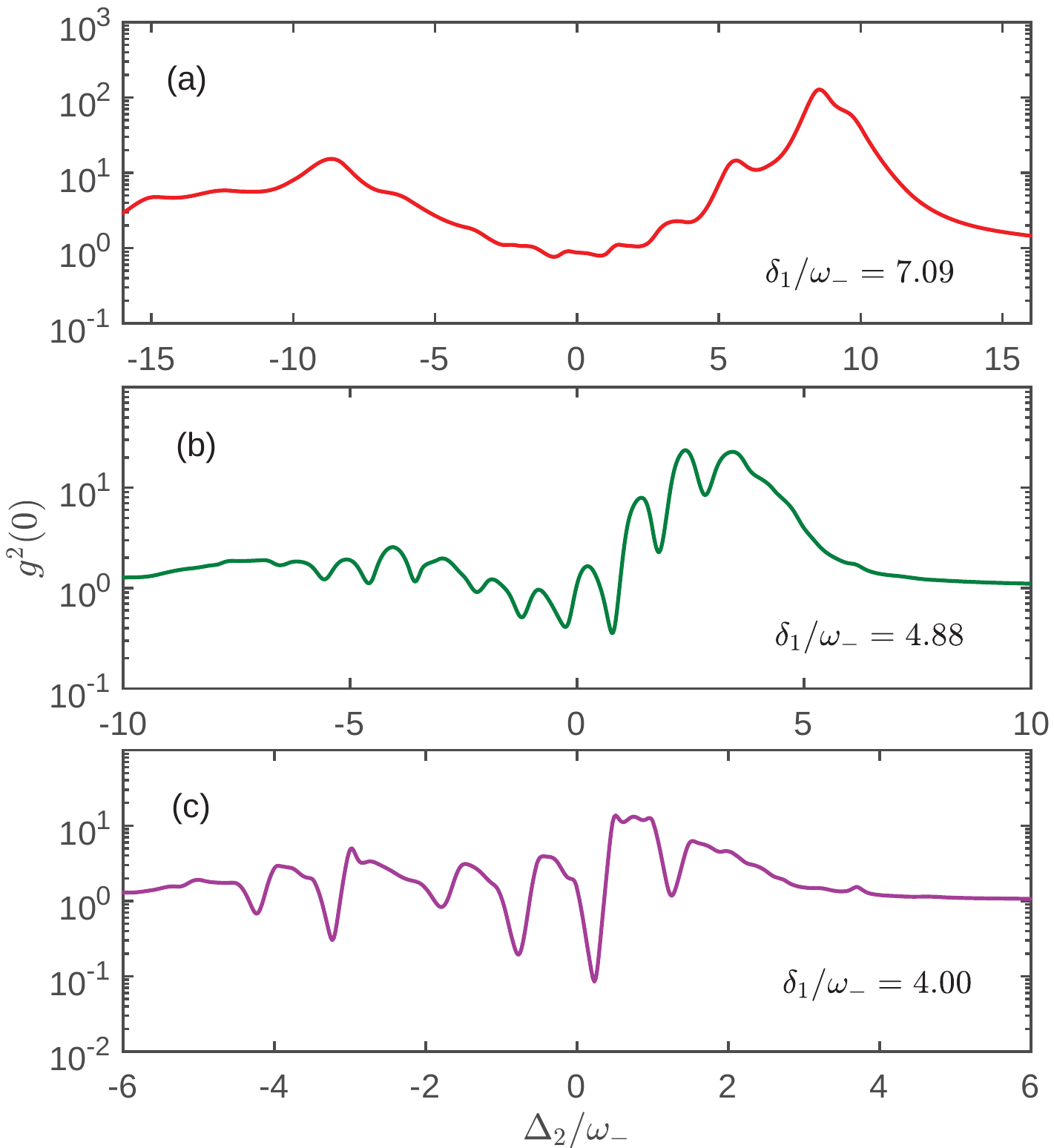}}
\caption{(Color online) The equal-time second-order correlation function $g^{2}(0)$ versus the driving detuning $\Delta_{2}/\omega_{-}$ for various values of the detuning $\delta_{1}/\omega_{-}$ for (a) $\delta_{1}/\omega_{-}=7.09$, $g_{1}/\omega_{-}=4.43\times10^{-2}$, $g_{2}/\omega_{-}=1.33\times10^{-2}$, (b) $\delta_{1}/\omega_{-}=4.88$, $g_{1}/\omega_{-}=2.80\times10^{-2}$, $g_{2}/\omega_{-}=8.37\times10^{-3}$, (c) $\delta_{1}/\omega_{-}=4.00$, $g_{1}/\omega_{-}=2.00\times10^{-2}$, $g_{2}/\omega_{-}=6.00\times10^{-3}$.}
\label{fig9}
\end{figure}
\begin{eqnarray}
\frac{d\rho}{dt}&=&-i[H_{\rm OMS},\rho]+\kappa \mathcal D[a_{2}]\rho+\kappa_{-}{\bar{n}}_{-}\mathcal D[B_{-}^{\dagger}]\rho  \nonumber  \\
&&+\kappa_{-}({\bar{n}}_{-}+1)\mathcal D[B_{-}]\rho
+\kappa_{+}{\bar{n}}_{+}\mathcal D[B_{+}^{\dagger}]\rho  \nonumber  \\
&&+\kappa_{+}({\bar{n}}_{+}+1)\mathcal D[B_{+}].
\end{eqnarray}
\begin{figure}[htb]
\centerline{\includegraphics[width=7.8cm]{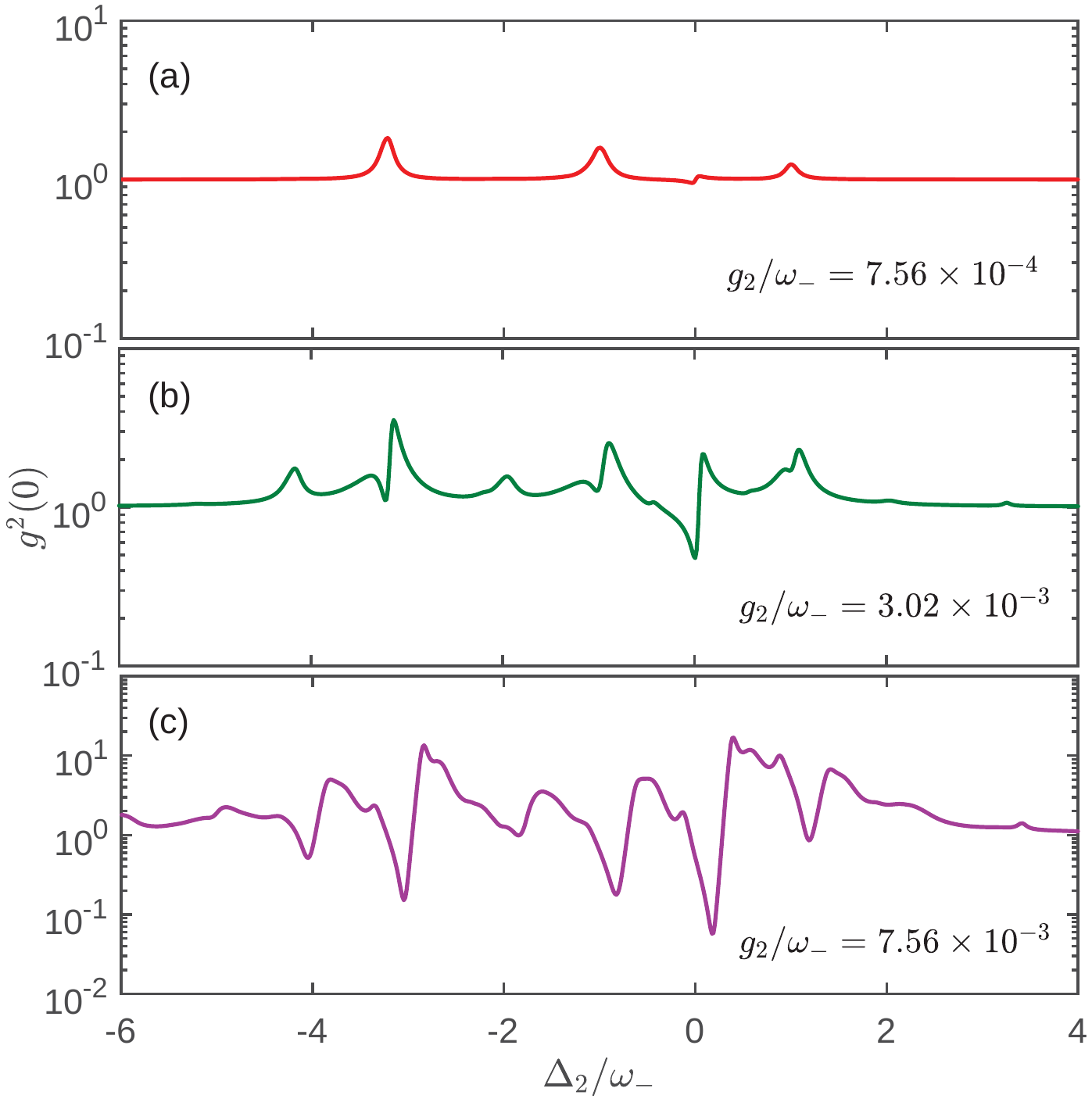}}
\caption{(Color online) The equal-time second-order correlation function $g^{2}(0)$ versus the driving detuning $\Delta_{2}/\omega_{-}$ for various values of the quadratic coupling strength $g_{2}/\omega_{-}$. Parameters are $\beta=25, g_{1}/\omega_{-}=1.51\times10^{-2}, \delta_{1}/\omega_{-}=3.78$ and (a) $ g_{2}/\omega_{-}=7.56 \times10^{-4}$, (b) $ g_{2}/\omega_{-}=3.02 \times 10^{-3}$, (c) $g_{2}/\omega_{-}=7.56 \times 10^{-3}$.}
\label{fig10}
\end{figure}Here, $\mathcal D[o]\rho=o\rho o^{\dagger}-(o^{\dagger}o\rho +\rho o^{\dagger}o)/2$ is the standard Lindblad superoperator for the damping of the polariton modes. The effective polariton damping rates $\kappa_{\pm}$ read
\begin{subequations}
\begin{align}
\kappa_{-}=\gamma\frac{\omega_{m}\rm sin^{2} \theta}{\omega_{-}}+\kappa\rm cos^{2} \theta , \\
\kappa_{+}=\gamma\frac{\omega_{m}\rm cos ^{2}\theta}{\omega_{+}}+\kappa\rm sin^{2}\theta  ,
\end{align}
\end{subequations}
where $\kappa$ and $\gamma$ are the initial cavity and mechanical damping rate, respectively. As one can see in Fig.\,\ref{fig6}, $\kappa_{+} $ decreases with increasing $\Delta_{1}/\omega_{m}$ while $\kappa_{-}$ becomes larger with increasing $\Delta_{1}/\omega_{m}$, but all $\kappa_{\pm}$ are smaller than initial cavity damping rate $\kappa$. Moreover, $\kappa_{\pm}$ have an inverse tendency with the growth of $G_{1}/\omega_{m}$. In the region discussed in this paper, the corresponding damping rates of subsystem can be suppressed effectively. We assume the mechanical bath to be at temperature $T_{\rm M} =0$, then the effective thermal occupancies for modes $B_{\pm}$ can be written as

\begin{subequations}
\begin{align}
\bar{n}_{-}=\frac{\kappa \rm cos ^{2}\theta}{4\kappa_{-}\Delta_{1}\omega_{-}}(\Delta_{1}-\omega_{-})^{2},\\
\bar{n}_{+}=\frac{\kappa \rm sin ^{2}\theta}{4\kappa_{+}\Delta_{1}\omega_{+}}(\Delta_{1}-\omega_{+})^{2}.
\end{align}
\end{subequations}
Here $\bar{n}_{\pm}$ originally come from the environments of the optical mode $a_1$ and mechanical mode $b$ [see the definition of $B_{\pm}$ shown in Eqs.\,(6)]. The detailed derivation of Eqs.\,(16) is given in the Appendix. In Fig.\,\ref{fig7}, we plot the relationship between effective bath thermal occupancies given in Eqs.\,(16) and variables $\Delta_{1}/\omega_{m}$ and $G/\omega_{m}$. It is shown that, in the considered region, $n_{\pm}$ are all less than 0.8, which ultimately ensures the probability to observe phonon sidebands and photon blockade effects in our proposal.

To demonstrate the phonon sideband, we calculate the steady-state excitation spectrum
\begin{align}
S(\Delta_{2})=\lim_{t \to \infty}\frac{\langle a_{2}^{\dagger}a_{2} \rangle(t)}{n_{0}},
\end{align}
where the resonant photon number $n_{0}=4\varepsilon^{2}/\kappa^{2}$. By using Eq.\,(17) with Hamiltonian $H_{\rm OMS}$, we plot the optical photon excitation spectrum $S(\Delta_{2})$ as a function of various detuning $\delta_{1}/\omega_{-}$ in Fig.\,\ref{fig8}. (Note that we plot Figs.\,\ref{fig8}, \ref{fig9} and \ref{fig10} using the Hamiltonian $H_{\rm OMS}$ and considering three modes, i.e., $a_{2}, B_{+}$ and $B_{-}$.)  It shows that phonon sideband appears in the originally weak coupling regime and also can be controlled by the detuning $\delta_{1}/\omega_{-}$. This clearly demonstrates the controllable optomechanical nonlinearity featured in our system, which could effectively enter into strong coupling regime, i.e., $g_{\pm}>\kappa_{\pm}$.

To demonstrate the photon blockade effect, we calculate the equal-time second-order correlation function in a steady state
\begin{align}
g^{(2)}(0)=\lim_{t\to 0}\frac{\langle a_{2}^{\dagger}a_{2}^{\dagger}a_{2}a_{2}\rangle(t)}{\langle a_{2}^{\dagger}a_{2}\rangle^{2}(t)}.
\end{align}
By numerically solving Eq.\,(18) with Hamiltonian $H_{\rm OMS}$, accordingly, the second-order correlation function $g^{2}(0)$ of the optomechanical system can be calculated.
In Figs.\,\ref{fig9} and \ref{fig10}, we illustrate the dependence of $g^{2}(0)$ on the weak probe field detuning $\Delta_{2}/\omega_{-}$ for different driving detuning $\delta_{1}/\omega_{-}$ and initial quadratic coupling strength $g_{2}/\omega_{-}$.

Figure \ref{fig9}(a) shows that there is no photon blockade ($g^{2}(0)\ll1$) for large detuning $\delta_{1}/\omega_{-}$ since large detuning causes weak nonlinearity and large effective bath thermal occupancies $n_{\pm}$. With the decrease of $\delta_{1}/\omega_{-}$, we observe a series of bunching (peaks) and untibunching (dips) resonances with the change of $\Delta_{2}/\omega_{-}$. This indicates that the nonlinear effect is enhanced extensively and the noise of the optomechanical system is suppressed effectively along with the decrease of driving detuning $\delta_{1}/\omega_{-}$. When $\delta_{1}/\omega_{-} \leqslant4.00$, the photon blockade effect appears in the certain values of $\Delta_2/\omega_{-}$ due to the system effectively reaching into the strong-coupling regime [see Fig.\,\ref{fig9}(c)]. Again, this results also demonstrate the controllable optomechanical nonlinearity can be realized in our proposal. In Fig.\,\ref{fig10}, we plot the second-order equal time correlation function $g^{2}(0)$ as a function of $\Delta_{2}/\omega_{-}$ when the initial quadratic coupling strength $g_{2}$ takes various values.  It is also clearly shown that the realization of photon blockade in the weak coupling regime based on our proposal. In other words, the strong optomechanical nonlinearity could be realized in the weakly-coupled OMS.
\section{discussions}\label{sectionV}
In this section, we discuss the experimental feasibility of our proposal. To obtain strong linearized interaction between the optical mode $a_1$ and mechanical mode $b$ (i.e., $G_1=\alpha_1g_1\sim0.1\omega_m$), a large optomechanical interaction strength $g_1/\omega_m=10^{-2}$ and cavity photon number $\langle a_1^{\dagger}a_1\rangle=\alpha_1^2=2500$ have been chosen here. Until now, it is still challenging to realize $g_1/\omega_m=10^{-2}$ in the normal optomechanical system~\cite{Aspelmeyer2014}. In principle, this problem can be solved by increasing the mean photon number $\langle a_1^{\dagger}a_1\rangle$ with applying a strong driving laser. As shown in Ref.\,\cite{Groblacher2009}, the cavity photon number up to $10^{10}$ has been realized experimentally, which manifests that the restriction of the parameter condition $g_1/\omega_m$ of our proposal can be relaxed largely theoretically.

Moreover, a relatively strong optomechanical coupling strength $g_1/\omega_{m}\sim10^{-2}$ has been obtained in the optomechanical crystal\,\cite{Eichenfield2009,Chan2009} or ultracold atomic optomechanical system\,\cite{Murch2009}. Recently, many potential theoretical schemes have been proposed to enhance the single-photon optomechanical coupling strength by utilizing the Josephson effect in superconducting circuits~\cite{Heikkila2014}, the squeezing effect\,\cite{Lv20151,Lemonde2016} and so on. In a short summary, it might be still challenging to implement our proposal completely with current available experimental technology. We hope it will be realized in the further experiments along with the progress of cavity optomechanics.
\section{conclusion}\label{sectionVI}

In conclusion, we presented a method to obtain controllable optomechanical nonlinearity in an OMS including both the linear and quadratic optomechanical coupling. By applying a strong driving laser into a cavity, we have shown that
the optomechanical coupling could be enhanced enormously without loss of the nonlinearity, i.e., from weak coupling regime to an effective strong radiation-pressure coupling regime. To demonstrate the controllable optomechanical nonlinearity, we numerically calculated cavity excitation spectrum and second-order correlation function, and presented the appearances of phonon sidebands and photon blockade effects in the originally weak coupling regime. Our results shown that, in the dual-coupling optomechanical system, one can easily enhance the optomechanical nonlinearity by applying a strong driving laser. This study provides a promising route to reach the strong nonlinear regime of an OMS with available technology, and has potential applications in modern quantum science.

\begin{acknowledgments}
This work is supported by the National Key Research and Development Program of China Grant No.2016YFA0301203, the National Science Foundation of China (Grant Nos. 11374116, 11574104 and 11375067).
\end{acknowledgments}
\appendix
\section{Derivation of the effective thermal occupancies}
In this appendix, we show the detailed discussion on the generation  of the effective thermal occupancies $\bar{n}_{\pm}$. Here $\bar{n}_{\pm}$ are the effective thermal occupancies of polariton modes $B_{+}$ and $B_{-}$. The initial master equation with Hamiltonian $H_{\rm eff}$ can be written as,
\begin{align}
\frac{d\rho}{dt}=&-i[H_{\rm eff},\rho(t)]+\kappa\mathcal D[a_{1}]\rho+\kappa\mathcal D[a_{2}]\rho \nonumber  \\
&+(\bar{n}+1)\gamma\mathcal D[b]\rho+\bar{n}\gamma\mathcal D[b^{\dagger}]\rho,
\end{align}
here, $\mathcal D[o]\rho=o\rho o^{\dagger}-(o^{\dagger}o\rho +\rho o^{\dagger}o)/2 $ is the standard Lindblad superoperator. $\bar{n}=\bar{n}_{B}[\omega_{m},T_{M}]=[{\rm exp}(\omega_{m}/k_{B}T_{M})-1]^{-1}$ is the mean number of the mechanical mode inside the bath and $k_{B}$ is Boltzmann constant and the mechanical bath is at temperature $T_{M}$.
Applying the inverse of Eqs.\,(6a-6d) into Eq.\,(A1), then we can obtain the effective master equation [see Eq.\,(14)]. Accordingly, the effective polariton damping rates are\,\cite{Lemonde2015}
\begin{subequations}
\begin{align}
\kappa_{-}=&\gamma(E_{+}+E_{-})^{2}+\kappa(C_{+}^{2}-C_{-}^{2}) \nonumber \\ =&\gamma\frac{\omega_{m}{\rm sin}^{2}\theta}{\omega_{-}}+\kappa {\rm cos}^{2}\theta ,  \\
\kappa_{+}=&\gamma(F_{+}+F_{-})^{2}+\kappa(D_{+}^{2}-D_{-}^{2})\nonumber \\ =&\gamma\frac{\omega_{m}{\rm cos}^{2}\theta}{\omega_{+}}+\kappa {\rm sin}^{2}\theta ,
\end{align}
\end{subequations}
where $C, D, E, F$ are matrix factors of Eqs.\,(5). Specifically, when the initial modes $a_{1}$ and $b$ transform into polariton modes $B_{+}$ and $B_-$, the interaction between cavity and bath generates nonconserving terms and these nonconserving terms lead to polariton heating. The thermal occupancies read,
\begin{subequations}
\begin{align}
\bar{n}_{-}&=\frac{1}{\kappa_{-}}[\gamma(E_{+}+E_{-})^{2}\bar{n}_{B}[\omega_{-},T_{M}]+\kappa C_{-}^{2}]\nonumber\\&=\frac{\gamma \omega_{m}{\rm sin}^{2}\theta}{\kappa_{-}\omega_{-}}\bar{n}_{B}[\omega_{-},T_{M}]+\frac{\kappa {\rm cos}^{2}\theta}{4\kappa_{-}\Delta_{1}\omega_{-}}(\Delta_{1}-\omega_{-})^{2} , \\
\bar{n}_{+}&=\frac{1}{\kappa_{+}}[\gamma(F_{+}+F_{-})^{2}\bar{n}_{B}[\omega_{+},T_{M}]+\kappa D_{-}^{2}]\nonumber\\&=\frac{\gamma \omega_{m}{\rm cos}^{2}\theta}{\kappa_{+}\omega_{+}}\bar{n}_{B}[\omega_{+},T_{M}]+\frac{\kappa {\rm sin}^{2}\theta}{4\kappa_{+}\Delta_{1}\omega_{+}}(\Delta_{1}-\omega_{+})^{2}.
\end{align}
\end{subequations}
We choose the temperature of the mechanical bath to be zero, i.e., $\bar{n}_{B}[\omega_{-},T_{M}]= \bar{n}_{B}[\omega_{+},T_{M}]=0$, then we obtain  the effective thermal occupancy as follows:
\begin{subequations}
\begin{align}
\bar{n}_{-}=\frac{\kappa {\rm cos} ^{2}\theta}{4\kappa_{-}\Delta_{1}\omega_{-}}(\Delta_{1}-\omega_{-})^{2},\\
\bar{n}_{+}=\frac{\kappa {\rm sin} ^{2}\theta}{4\kappa_{+}\Delta_{1}\omega_{+}}(\Delta_{1}-\omega_{+})^{2}.
\end{align}
\end{subequations}

\end{document}